\documentclass{aa}  

\usepackage{graphicx}
\usepackage{subfigure}
\usepackage{lscape}
\usepackage{txfonts}
\begin{document}

   \title{Correlation between activity indicators:  H$\alpha$ and Ca II lines in M-dwarf stars}


   \author{R.V. Iba\~nez Bustos
\inst{1,2}\fnmsep\thanks{Contact e-mail: ribanez@iafe.uba.ar}
          \thanks{Based on data obtained at Complejo Astron\'omico El Leoncito, operated under the agreement between the Consejo Nacional de Investigaciones Cient\'\i ficas y T\'ecnicas de la Rep\'ublica Argentina and the National Universities of La Plata, C\'ordoba and San Juan.}
          \and
          A.P. Buccino\inst{1,3}
          \and
          M. Flores\inst{4,5}
          \and
          C. F. Martinez\inst{1,3}
          \and
          P.J.D. Mauas\inst{1,3}
          }

   \institute{Instituto de Astronom\'ia y F\'isica del Espacio (CONICET-UBA), C.C. 67 Sucursal 28, C1428EHA-Buenos Aires, Argentina.\\
        \and
             Departamento de F\'isica, Facultad de Ingeniería, Universidad de Buenos Aires, Buenos Aires, Argentina.\\
        \and
             Departamento de F\'isica, Facultad de Ciencias Exactas y Naturales, Universidad de Buenos Aires, Buenos Aires, Argentina.\\
        \and
             Instituto de Ciencias Astron\'omicas, de la Tierra y del Espacio (ICATE-CONICET), San Juan, Argentina.\\
        \and
            Facultad de Ciencias Exactas, F\'isicas y Naturales, Universidad Nacional de San Juan, San Juan, Argentina.
            }

   \date{}

 
  \abstract
   {Different approaches have been adopted to study both short- and long-term stellar magnetic activity, and although the mechanisms by which low-mass stars generate large-scale magnetic fields are not well understood, it is known that stellar rotation plays a key role.}
   {There are stars that show a cyclical behaviour in their activity studied on the blue side of the visible spectrum, which can be explained by solar dynamo or $\alpha\Omega$ dynamo models. However, when studying late-type dwarf stars, they become redder and it is necessary to implement other indicators to analyse their magnetic activity. In the present work, we perform a comparative study between the best-known activity indicators so far defined from the Ca II and H$\alpha$ lines to analyse M-dwarf stars.}
   {We studied a sample of 29 M stars with different chromospheric activity levels and spectral classes ranging from dM0 to dM6. To do so, we employed 1796 spectra from different instruments with a median time span of observations of 21 years. The spectra have a wide spectral range that allowed us to compute the chrosmospheric activity indicators based on Ca II and H$\alpha$. In addition, we complemented our data with photometric observations from the \textit{TESS} space mission for better stellar characterisation and short-term analysis.}
   {We obtained a good and significant correlation ($\rho = 0.91$) between the indexes defined from the two lines for the whole set of stars in the sample. However, we found that there is a deviation for faster rotators (with $P_{rot}$ < 4 days) and higher flare activity (at least one flare per day).
   For the individual analysis, we found that the indexes computed individually for each star correlate independently of the level of chromospheric emission and the rotation period.}
   {There is an overall positive correlation between Ca II and H$\alpha$ emission in dM stars, except during flare events. In particular, we found that low-energy high-frequency flares could be responsible for the deviation in the linear trend in fast-rotator M dwarfs. This implies that the rotation period could be a fundamental parameter to study the stellar activity and that the rotation could drive the magnetic dynamo in low-mass active stars. }


   \maketitle
%

\section{Introduction}

M dwarfs, with masses between 0.1 and 0.5 M$_\odot $, constitute $\sim75 \%$ of the stars in the solar neighbourhood.
Because of their low mass, M stars represent an important laboratory for detecting Earth-like planets orbiting around them.
On the other hand, they have a high occurrence rate of extrasolar planets orbiting in the habitable zone\footnote{The habitable zone is defined as the region in the equatorial plane of the star where the water in the planet, if any, can be found in its liquid state.} surrounding the star (\citealt{Bonfils13, DressingCharbonneau15}). 
However, several of these M stars present high chromospheric activity that can exceed the solar magnetic activity. 
They are usually called flare stars due to the high frequency of these high-energy transient events. 
Thus, an Earth-type planet orbiting an M star could be frequently affected by small short flares and eventually by long high-energy flares, a fact which could constrain the exoplanet's habitability \citep{Buccino07, Vida17} or even play an important role in the atmospheric chemistry of an orbiting planet (\citealt{Miguel15}).

On the other hand, the phenomena associated with stellar activity can also be studied from several spectral lines that are sensitive to chromospheric activity (e.g. Mg \scriptsize{II}\normalsize\-, Ca \scriptsize{II}\normalsize\-, Na \scriptsize{II}\normalsize\-, and H$\alpha$).
The mean integrated line-core fluxes are used as indicators of the total activity level of the stars, and their variability with time is related to stellar rotation and the evolution of the stellar active regions. They provide information on the different regions at a different height of the stellar atmosphere. 

In 2007, \citeauthor{Livingston07} confirmed the correlation between the H\&K lines of the Ca \scriptsize{II}\normalsize\- and H$\alpha$ emissions during the 11-year activity cycle of the Sun.
Several authors have suggested that such a correlation is also present in other stars (\citealt{Giampapa89, Pasquini91, Montes95}).
From there on, several works were published focussed on the analysis of the relation and behaviour of both lines either in the Sun or in FGK and even in M stars (\citealt{Cincunegui07b, Meunier09, Santos10, GomesdaSilva11, GomesdaSilva14, Flores18, Meunier22}).

However, when \cite{Cincunegui07b} studied the relation between simultaneous measurements of both Ca  \scriptsize{II}\normalsize\- and H$\alpha$  fluxes  in a sample of 109 FGK and M stars observable from the southern hemisphere, they found a large scatter in the correlation coefficient when analysing each star individually, ranging from very strong positive to negative correlations.
They also suggested that the mean flux values in the Ca \scriptsize{II}\normalsize\- and H$\alpha$ lines are correlated due to the effect of stellar colour on both fluxes.
\cite{Meunier09} studied the contribution of the plages and filaments to the emission in the Ca \scriptsize{II}\normalsize\- and H$\alpha$ lines during a solar cycle. 
In their work they found that the {plages} contribute to an increase in the emission of both fluxes, while the filaments pronounce the absorption only in H$\alpha$. 
Furthermore, they reported that the contribution of filaments to H$\alpha$ may be responsible for the decrease in the correlation coefficient between the two fluxes as a function of their spatial distribution and contrast.
They also noted that at higher activity levels (e.g. during the maximum of the cycle),
the filling factor of the filament saturates  and the correlation between the two fluxes increases \citep{Meunier22}.

Other factors contributing to a decrease in the measured correlation may be the time interval of the observations, the phase of the cycle in which they are measured, and the stellar inclination angle. 
In 2009, \citeauthor{Walkowicz09} observed a strong positive correlation between individual observations of the Ca \scriptsize{II}\normalsize\- and H$\alpha$ lines in most of the active dM3e stars in their sample.
\cite{Santos10} studied the long-term activity of eight FGK stars using the Mount Wilson and H$\alpha$ indexes and found a general long-term correlation between the two indicators. 
However, their sample was not large enough for statistical significance.
\cite{GomesdaSilva11} extended the comparison between these two activity-sensitive lines to early M dwarfs. 
Similar to \cite{Cincunegui07b}, they detected a wide variety of correlation coefficients, including anti-correlations for the less active stars in their sample. 
However, all of the most active stars were positively correlated. 
They also found indications that the H$\alpha$ index was following an `anti-cycle' relative to its $S$ index in some cases, that is to say the maxima and minima measured in the two indexes were anti-correlated.
Recently, \cite{Gomesdasilva22} studied the correlation between both lines for FGK dwarfs stars observed with HARPS (High Accuracy Radial velocity Planet Searcher).
They found that the anti-correlation in these stars could become a positive relation between indexes when they changed for a narrow bandwith on H$\alpha$.

On the other hand, several studies suggest that the H$\alpha$ emission line is strongly affected by the magnetic field and rotation period (e.g. \citealt{Reiners12, Newton17}), which are fundamental keys to the  flare mechanism \citep{Yang17}.
With the aim of extending the results of the aforementioned works, in the present study we analyse the correlation between simultaneous measurements of the Ca \scriptsize{II}\normalsize\- and H$\alpha$ line-core fluxes for 29 M stars with different chromospheric activity levels and spectral classes ranging from dM0 to dM6.


 \section{The H$\alpha$ line}

Although Ca \scriptsize{II}\normalsize\- resonance lines have been the most widely used chromospheric indicators for stellar activity studies, they have a number of disadvantages to study late-type dwarf stars. 
First, late K and M stars are progressively redder than solar-type stars and, therefore,  the  blue Ca \scriptsize{II}\normalsize\- lines become less intense than other redder features in the visible spectrum.
Second, red dwarfs become much fainter as the spectral class increases, making them even more difficult to observe.
As a consequence, the Ca \scriptsize{II}\normalsize\- lines are not the most suitable for observational studies of cooler stars since the signal-to-noise ratio for these lines is usually very low and therefore they require long exposure times to get a reliable observation.

Although the formation of the H$\alpha$ line is dominated by photoionisation  in most solar-type dwarfs, towards late-type stars, decreasing photospheric radiation temperatures result in a more significant contribution of collisional processes to the source function\footnote{The source function is defined as the ratio between the energy emitted and lost when passing through the stellar atmosphere.} and eventually the  H$\alpha$ line core begins emission for the most active dKe-dMe stars (\citealt{Mauas94, Mauas96, Mauas97, Fontenla16}).  
These  latter stars  are  called  active  `dMe',  while  M  dwarfs  with  H$\alpha$ in absorption  are  usually  called  inactive.  
M  dwarfs  with  neither emission nor absorption in the H$\alpha$ line may either be very inactive or moderately active \citep{Walkowicz09}.
On the other hand, the spectral location of this line represents an observational advantage for late-type stars since the integration times required to obtain an adequate signal-to-noise ratio are much lower than those required when observing Ca \scriptsize{II}\normalsize\- lines.
However, several studies have questioned the use of the H$\alpha$ line as an indicator of long-term chromospheric activity \citep{Hawley03,Cincunegui07b,GomesdaSilva11,GomesdaSilva14,Buccino14,Flores18}.


\section{Observations and data reduction}

The HK$\alpha$ Project has been operating since 1999 with the main purpose being to study the long-term chromospheric activity of cool main sequence stars observable from the southern hemisphere. 
In  this  programme, we systematically observed late-type stars from dF5 to dM5.5 with the 2.15 m Jorge Sahade telescope at the {Complejo Astronómico El Leoncito}   Observatory (CASLEO), which is located at 2552 m above sea level in the Argentinian Andes.
The medium-resolution echelle spectra ($R\approx 13.000$) were obtained with the REOSC\footnote{\textsf{http://www.casleo.gov.ar/instrumental/js-reosc.php}} spectrograph, covering a wavelength range from 386 to 669 nm.
We calibrated all our echelle spectra in flux using IRAF\footnote{The Image Reduction and Analysis Facility (IRAF) is distributed by the Association of Universities for Research in Astronomy (AURA),  Inc., under contract to the National Science Foundation} routines and following the procedure described in \cite{Cincunegui04}.
For this study, we employed 368 observations distributed between 2000 and 2019. 
Each observation consists of two consecutive spectra with the same exposure time, which were combined to eliminate cosmic rays and to reduce noise (see details in \citealt{Cincunegui04}). 

We complemented our data with public observations obtained with several spectrographs. 
We used 1061 echelle spectra observed by HARPS, which is mounted at the 3.6 m telescope (\textit{R} $\sim$ 115.000) at La Silla Observatory (LSO, Chile), and distributed during the interval 2003 - 2018.\ In our study,
99 FEROS spectra were available. 
This spectrograph is placed on the 2.2 m telescope in LSO and has a resolution of $R \sim 48.000$.
Furthermore, 55 spectra were taken between 2002 and 2016 with UVES, which is attached to the Unit Telescope 2 (UT2) of the Very Large Telescope (VLT) ($R \sim 80.000$) at Paranal Observatory.
In addition, 23 medium-resolution spectra (\textit{R} $\sim$ 8.900) in the UVB wavelength range (300 - 559.5 nm) were obtained during the range 2010 - 2018 with the X-SHOOTER spectrograph, mounted at the UT2 Cassegrain focus also at the VLT. 
Finally, we employed 190 spectra from the HIRES spectrograph mounted at the Keck-I telescope which were observed between 1996 and 2017.

HARPS and FEROS spectra were automatically processed by their respective pipelines\footnote{\textsf{http://www.eso.org/sci/facilities/lasilla/instruments/harps.html}}$^,$\footnote{\textsf{http://www.eso.org/sci/facilities/lasilla/instruments/feros.html}}, while UVES and XSHOOTER observations  were manually processed with the corresponding method\footnote{\textsf{http://www.eso.org/sci/facilities/paranal/instruments/uves.html}}$^,$\footnote{\textsf{http://www.eso.org/sci/facilities/paranal/instruments/xshooter.html}}. 
For this study we employed 1796 spectra with a wide spectral range, which allowed us to analyse the most common spectral lines simultaneously in the optical range used to study magnetic activity, Ca \scriptsize{II}\normalsize\- and H$\alpha$.


\section{Stellar sample}

\begin{figure*}[h!]
\centering
\includegraphics[width=0.9\textwidth]{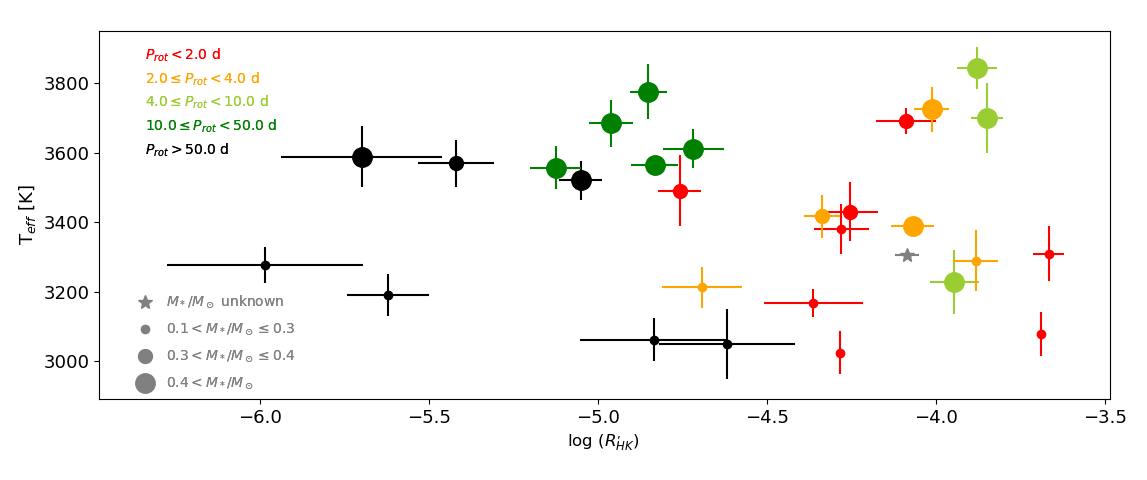}
\caption{M-dwarf sample studied in this work spans a range of $\sim$1000 K in effective temperature and $\sim$3 dex in activity levels as represented by the $\log R'_{HK}$ values. The relative mass of each star is represented by the size of the circles, with a range from 0.12 to 0.61$M_\odot$. The stellar rotation period is indicated by the colour of the circles, with a range from  1 to 145 days. The grey symbol represents a star with an unknown effective temperature; we show it as an active star in accordance with its $\log R'_{HK}$ value. This plot shows 29 stars for which the $\log R'_{HK}$ value is reported in this work. Meanwhile, the $T_{eff}$ and stellar masses were extracted from literature.}\label{sample}
\end{figure*}
With the main aim being to study the correlation between indexes from Ca \scriptsize{II}\normalsize\- and H$\alpha$ lines, we analysed a diverse sample of M dwarfs with different chromospheric activity levels.
All of these stars have been extensively observed during years even by the HK$\alpha$ Project sample and in the ESO and W. M. Keck observatories.
We included 16 active and 13 inactive stars, with a stellar spectral type from dM0e to dM6e for active ones, and dM1 - dM4.5 for the inactive dwarfs.
Their respective stellar rotation periods span between 1 day and 145 days.
Some of these rotation periods were detected in this work from their respective TESS/Kepler light curves and by employing the generalised Lomb-Scargle (GLS) periodogram (\citealt{Zechmeister09}).
Eleven stars on the sample are fully convective stars with masses $M_* / M_\odot < 0.3$, five stars are considered in the convective limit with stellar masses between $0.3 \leq M_* / M_\odot < 0.4$, and 12 stars were catalogued as partially convective stars with $M_* / M_\odot > 0.4$.
The stellar mass of only one star (Gl 1049) has not been reported in literature.
All aforementioned parameters are summarised in the graphic of Figure \ref{sample} and listed in Table \ref{tab_Halpha_resultados} with their respective references in literature.


\section{Chromospheric indexes' relationship}

\begin{figure}[h!]
\centering 
   \subfigure[\label{calcio_index}]{\includegraphics[width=9cm]{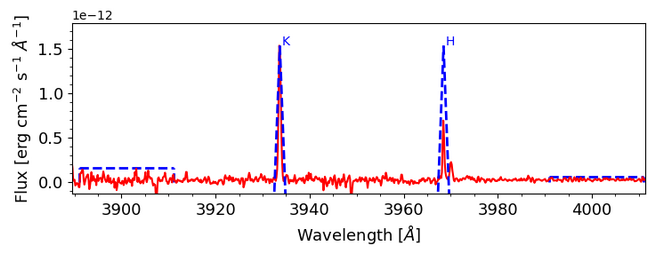}}
   \subfigure[\label{halpha_index}]{\includegraphics[width=9cm]{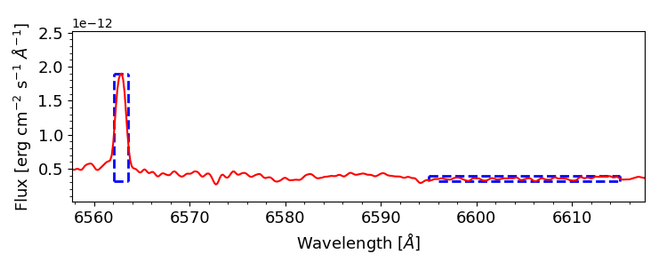}}
   \caption{Definition of the $S$ and $A$ indexes employed in our study.}
   \label{halpha_index}
\end{figure}

Stellar activity is usually characterised in Ca  \scriptsize{II}\normalsize\- lines by a dimensionless $S$ index, defined as the ratio between the chromospheric Ca \scriptsize{II}\normalsize\- H\&K line-core emissions, integrated with a triangular profile of 1.09 \AA\- full width at half maximum (FWHM), and the photospheric continuum fluxes integrated in two 20 \AA\- passbands centred at 3891 and 4001 \AA\-  (\citealt{Vaughan78}, see Fig. \ref{calcio_index}). 
From the $S$ index calculated for each spectrum of the stars in our sample, we obtained the chromospheric emission level $\log R'_{HK}$ following \cite{Astudillo17}: 

\begin{equation}
    R'_{HK} = R_{HK} - R_{phot} = K \sigma^{-1} 10^{-14} C_{cf} (S - S_{phot})
    \label{indice_logRHK}
,\end{equation}
where $R_{phot}$ and $S_{phot}$ are the photospheric contributions to $R$ and $S$, $C_{cf}$ is the bolometric factor estimated from the colour index of a given star\footnote{To estimate the bolometric factor, we employed both $(B-V)$ and $(V-K)$ colour indexes.}, $\sigma$ is the Stefan-Boltzmann constant, and $K$ is a conversion factor.

Similarly, we obtained the $A$ index defined in \cite{Cincunegui07b} as follows:

\begin{equation}
    A = \frac{f_{H\alpha}}{f_{cont}}
    \label{indiceA}
,\end{equation}
where $f_{H\alpha}$ is the average flux or number of counts measured in a 1.5 \AA\- window centred at the  H$\alpha$ line  (6562.8 \AA) and $f_{cont}$ is the average flux or number of counts in the  continuum nearby, centred at 6605 \AA\- and with a width of 20 \AA\-   (see Figure \ref{halpha_index}).

\begin{figure}[h!]
   \includegraphics[width=0.45\textwidth]{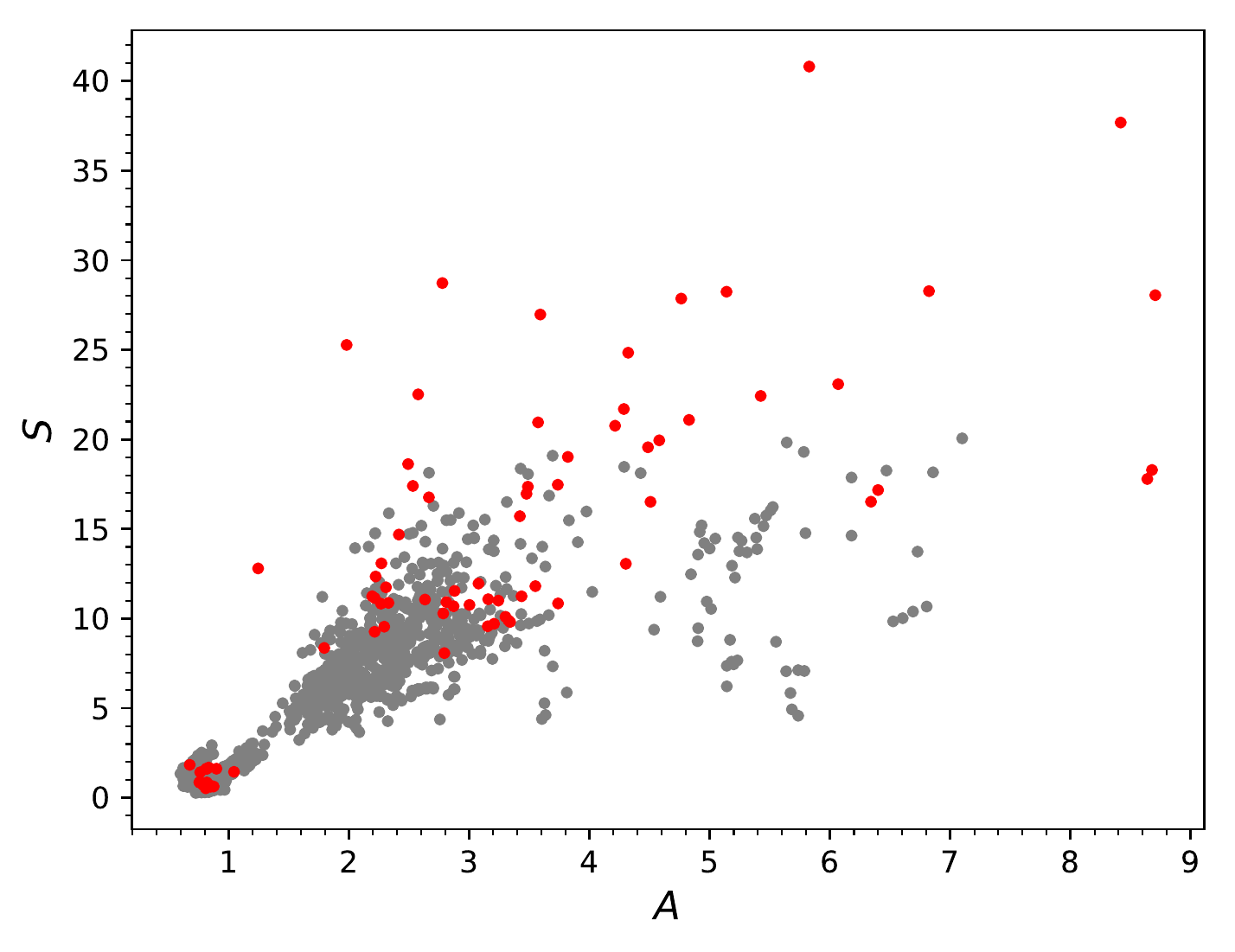}
   \caption{Mount Wilson $S$ indexes versus $A$ indexes for 1796 spectra used in our study. The flare-type transient events are highlighted in red.}
   \label{s-a_crudos}
\end{figure}

\begin{figure}[h!]
\centering
\includegraphics[width=9cm]{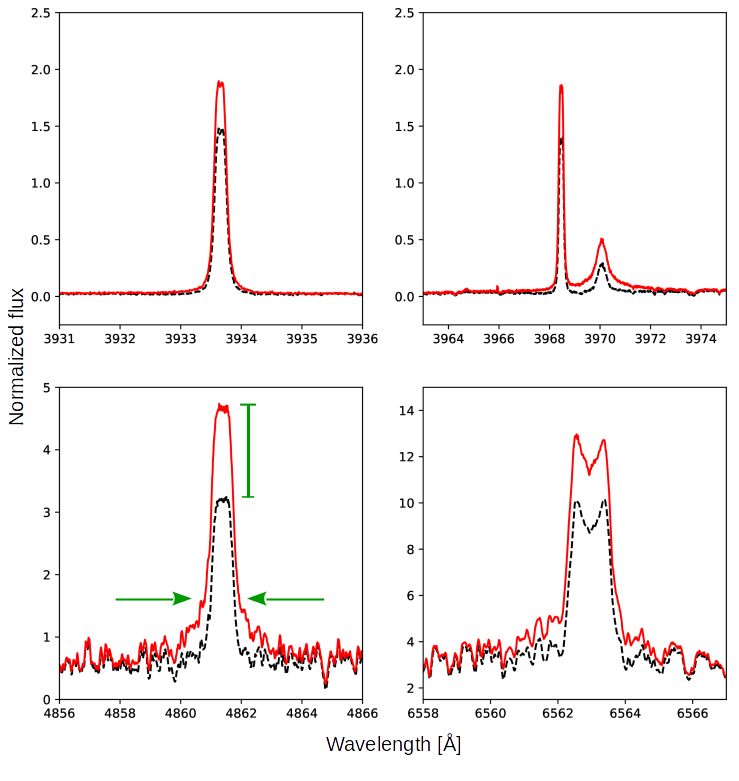}
\caption{Lines of Ca II H (top right) and K (top left), as well as H$\beta$ (bottom left) and H$\alpha$ (bottom right) for two HARPS spectra with HJD = 3833 (black dashed line) and HJD = 3834 (red line). During a flare we can see a broadening and intensification of the Balmer lines which allowed us to discriminate spectra contaminated by flare-type events. }\label{gl388_flares}
\end{figure}

From these adimensional indexes, we can extend the activity analysis including the public spectra not calibrated in flux that present simultaneous observations of the H\&K lines of Ca \scriptsize{II}\normalsize\- and H$\alpha$.
In Fig. \ref{s-a_crudos} we show the relation between the $S$ and $A$ indexes for the 1796 spectra.
In red we highlight the 79 transient flare-like events that we detected studying the line shapes, their core intensification, and the broadening of Balmer lines (especially the H$\beta$ line at 4861 \AA) as we show in Fig. \ref{gl388_flares}. 
We found that there is a general linear relationship for the grey points with $\rho = 0.87$, where the flares show a random behaviour between indexes. 
That is, high levels of activity in the $S$ index do not necessarily imply high levels of activity in the $A$ index.
This means that the difference between a quiet state of a star and a transient event may not be simultaneously reflected in the flux of Ca \scriptsize{II}\normalsize\- and H$\alpha$ lines and, therefore, we need to evaluate other chromospheric indicators in order to detect such events. 

\begin{figure}[h!]
\centering
   \subfigure[\label{s-a_bv}]{\includegraphics[width=0.45\textwidth]{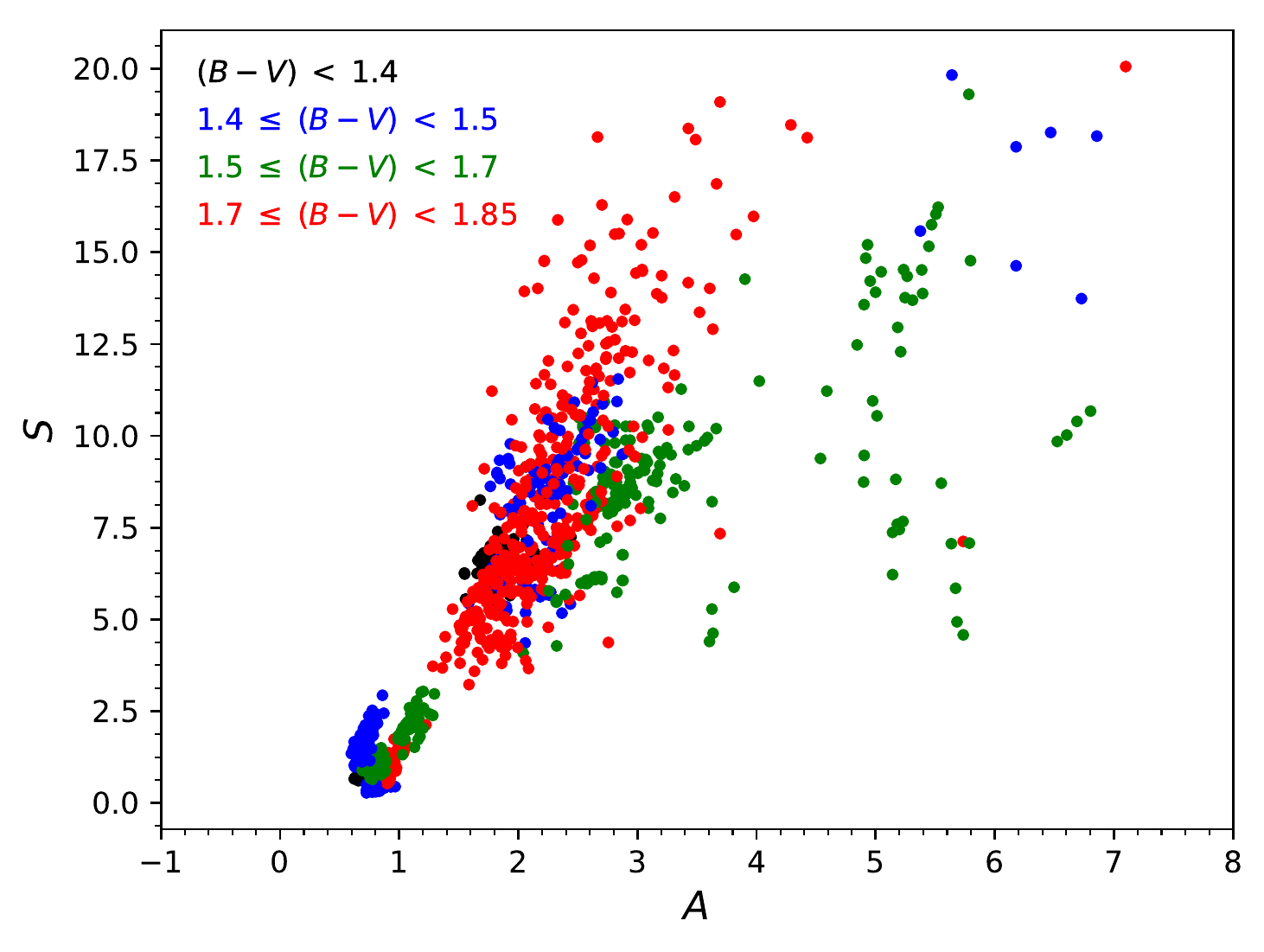}}
   \subfigure[\label{s-a_logRHK}]{\includegraphics[width=0.45\textwidth]{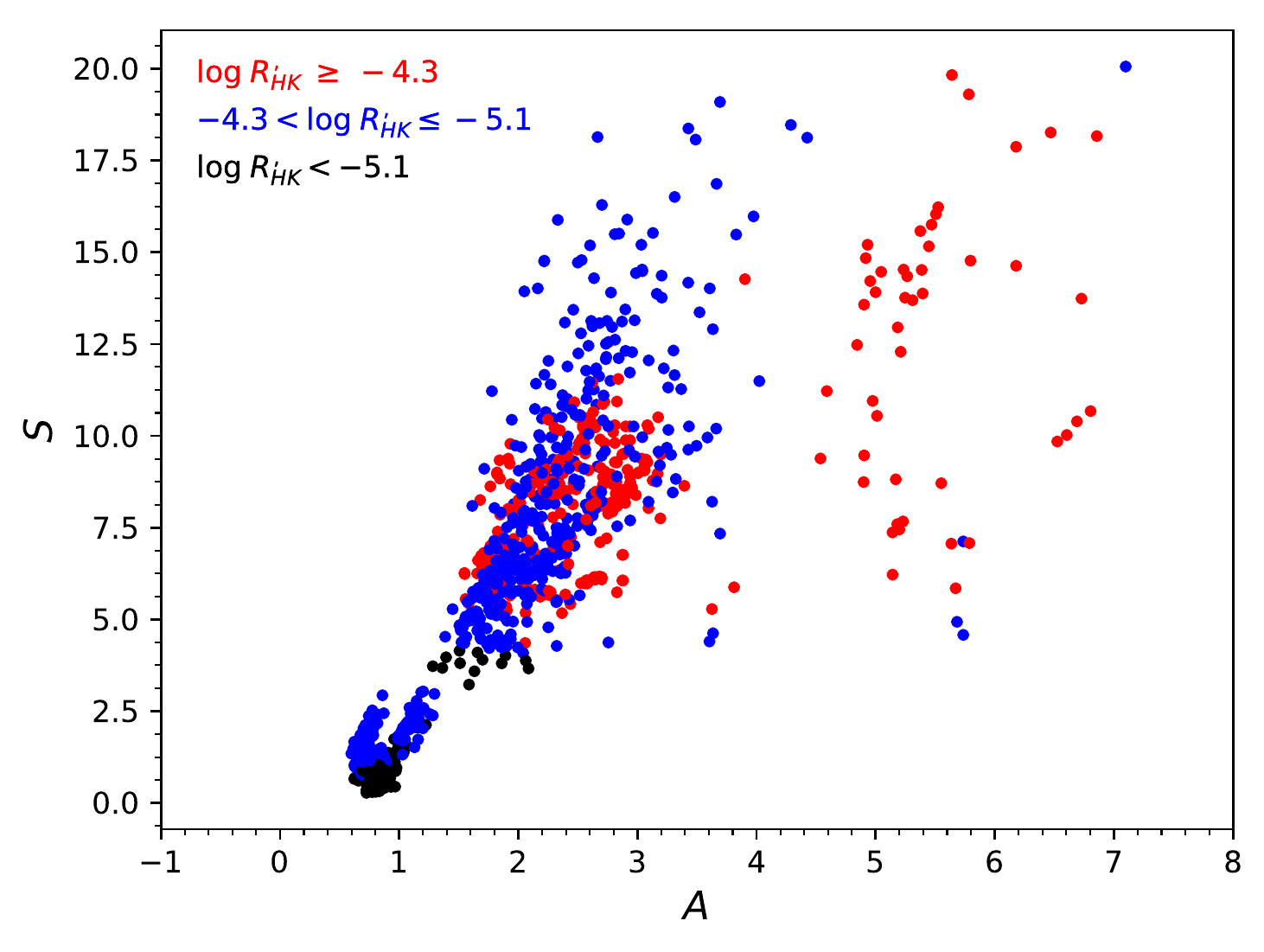}}
   \subfigure[\label{s-a_prot}]{\includegraphics[width=0.45\textwidth]{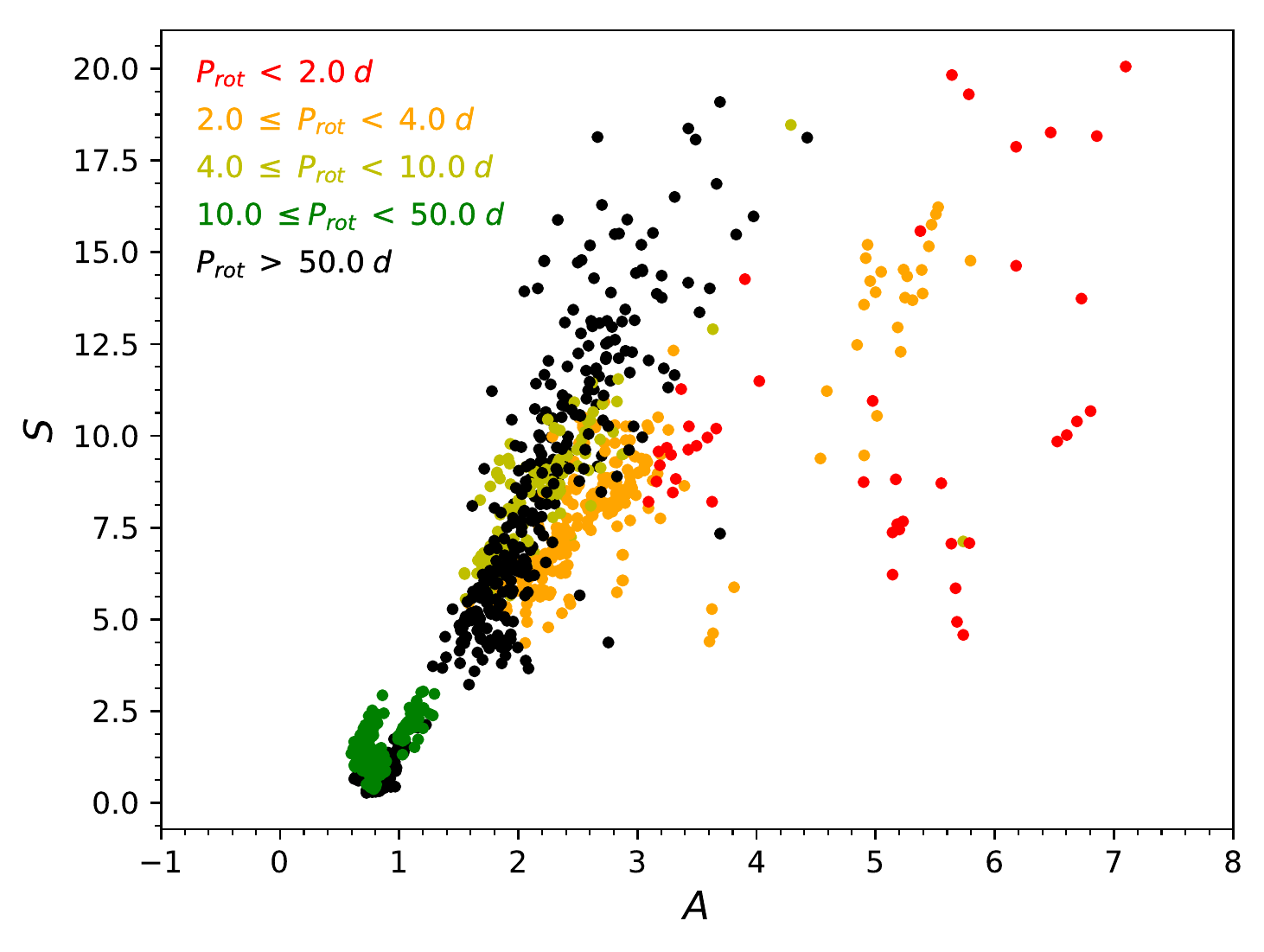}}
   \caption{$S$ index as a function of the  $A$ index without considering the flares and discriminating by colour (a), activity (b), and rotation period (c). 
   }
   \label{s-a_param}
\end{figure}

In Figs. \ref{s-a_bv}-\ref{s-a_prot} we show the $S$ index  as a function of the $A$ index without considering the flares and discriminating by colour, activity, and rotation period.
In Fig. \ref{s-a_bv} we cannot only notice an overall linear trend among the indexes, but also a departure from this trend for some of the redder stars. 
In Fig. \ref{s-a_logRHK} we observe that this deviation corresponds to those stars with higher chromospheric emission levels and within the $\pm3\sigma$ of the saturation level reported by \cite{Astudillo17}.
While in Fig. \ref{s-a_prot} we see that there is a critical rotation period of about $\sim$4 days, and that stars with rotation periods shorter than this deviate from the linear trend.

\begin{figure}[h!]
\centering
    \includegraphics[width=0.45\textwidth]{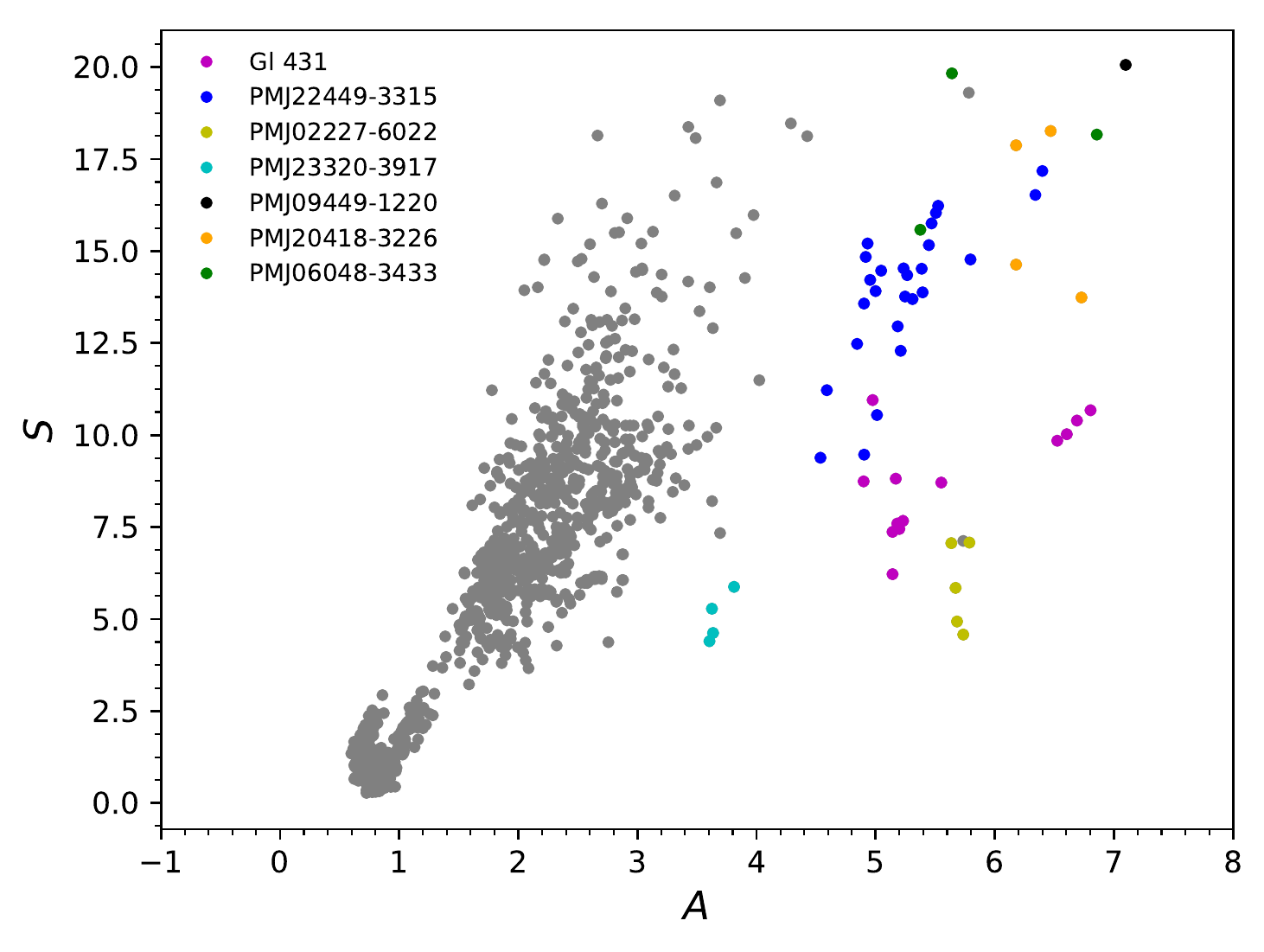}
    \caption{$S$ vs $A$ indexes. We identify stars that deviate from the linear behaviour between the indexes.}
    \label{s-a_estrellas}
\end{figure}

We identified those stars that deviate from the linear regime in Fig. \ref{s-a_estrellas}.
These M dwarfs present H$\alpha$ in emission, and they are fast rotators in the saturation regime of the $\log R'_{HK} - P_{rot}$ diagram within the $\pm3\sigma$ of the fit and with masses in the range (0.14 - 0.32) $M_\odot$.
On the other hand, we performed flare activity analysis for these stars following \cite{Ibanez20}  (see Appendix \ref{ap.flare}) and we found that they have a flare rate of at least one event per day.
\cite{Newton17} reported that rapidly rotating M dwarfs (aged less than 2 $\times10^9$ years) exhibit higher H$\alpha$ emission levels than slower rotators (aged more than 5 $\times10^9$ years).
More recently, \cite{Rodriguez20} reported that most flares show moderate to strong H$\alpha$ emission.
In agreement with \citeauthor{Newton17}, their results showed that those stars whose rotation periods are shorter show stronger H$\alpha$ emission than M dwarfs classified as non-flaring.
Our results are in agreement with both works.
However, they also indicate that in the faster rotators with $P_{rot}<4$ days, the strong emission in the upper chromosphere is not observed at lower altitudes, since the $S$ index remains within the ranges observed for the slower rotators.

\begin{figure}[htb!]
\centering
   \includegraphics[width=0.45\textwidth]{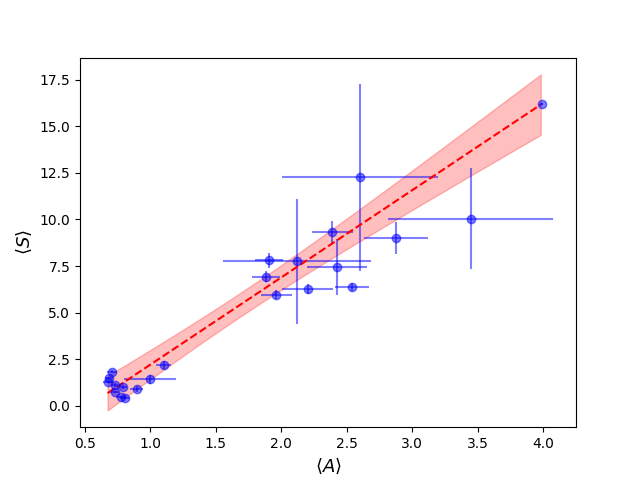}
   \caption{
   Averages of each individual star weighted by their errors. With the red dashed line, we show the best fit for our averages and with the red shading the 95\% confidence level.
   }
   \label{s-a_prompond}
\end{figure}

If we discard the stars identified in Fig. \ref{s-a_estrellas}, and if we take the average indexes for each star weighted with their corresponding errors, we obtain the  fit 
\begin{equation}
\langle S\rangle=(4.68 \pm 0.09) \langle A \rangle - (2.50 \pm 0.20)
\end{equation}
with a correlation coefficient of $\rho = 0.77$ in Fig. \ref{s-a_prompond}.

\begin{figure*}[htb!]
\centering
\includegraphics[width=0.9\textwidth]{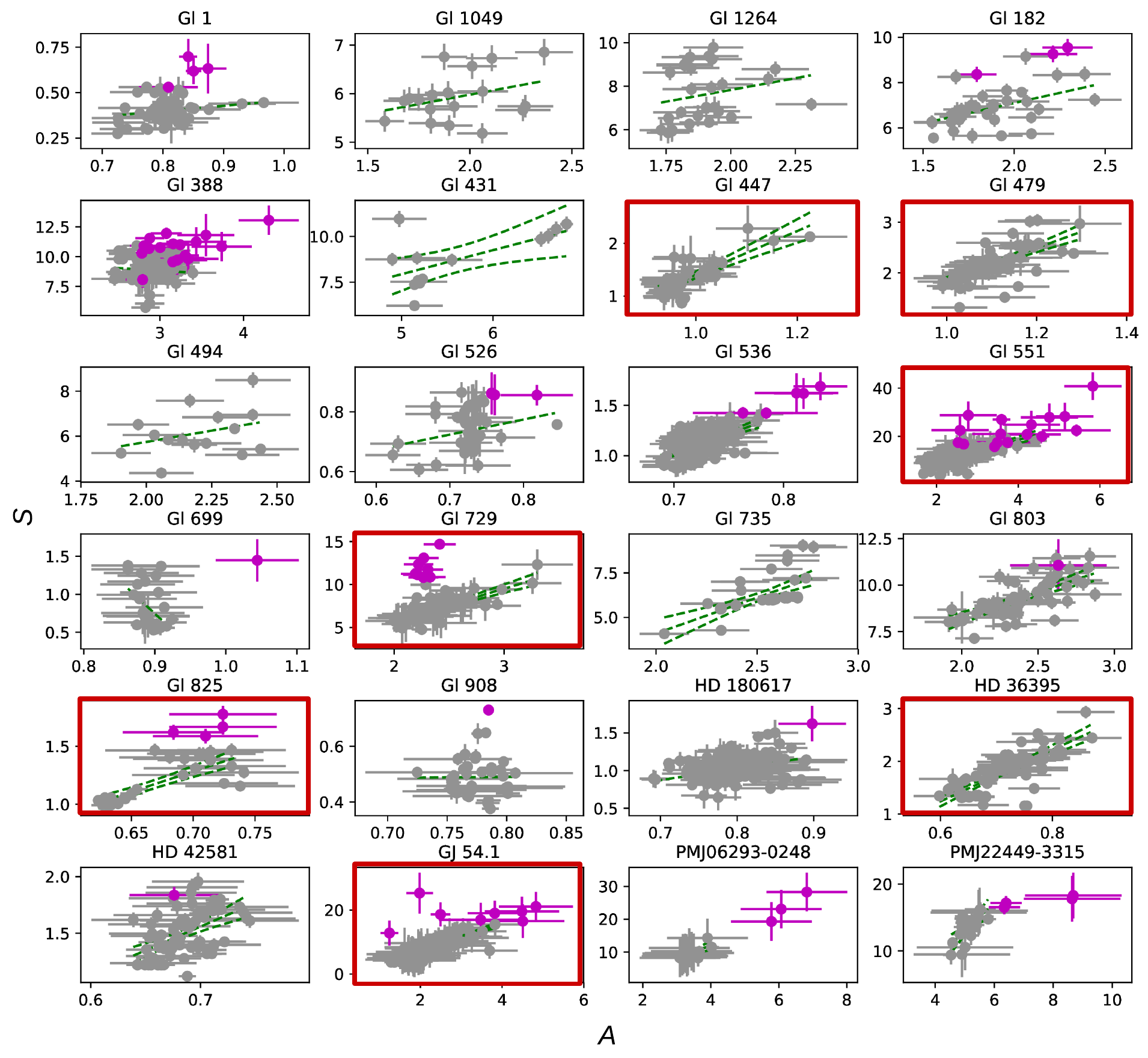}
\caption{
Relationship between the $S$ and $A$ indexes for each individual star. With the magenta circles, we show the detected flares for each star. With the green dashed line, we show the best fit line to the data without considering the flares. For those stars where the indexes correlate to a value of $\rho\geq 0.5$, we show the confidence level of 95\%. With red boxes we highlight those stars whose correlation coefficient is greater than 0.7.
}\label{s-a_indiv}
\end{figure*}

In Fig. \ref{s-a_indiv} we show simultaneous measurements of the $S$ index  as a function of the $A$ index for 24 of the 29 stars analysed in this work.
We discarded those stars that do not have enough observations (less than six observations) to study the correlation between indexes.

With magenta circles we represent the flare-type events.
For each star, flares do not only occur when both indexes adopt large values, but also for intermediate and random values of the indexes.

\begin{figure}[h!]
\centering
   \includegraphics[width=0.45\textwidth]{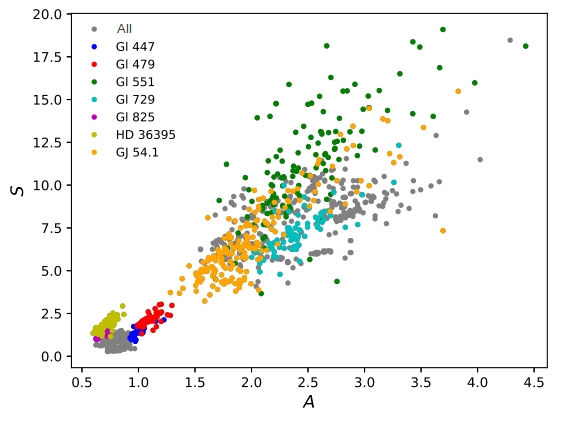}
       \caption{$S$ vs $A$ indexes for those stars in the sample with individual $\rho \geq 0.7$ in  Fig. \ref{s-a_indiv} }
   \label{s-a_indivCorr}
\end{figure}

In general, for most of the stars, the indexes are positively correlated or uncorrelated.
We performed a quantitative analysis using the \textsf{python} code provided by \cite{Figueira16}. We estimated the a posteriori probability distribution of the correlation coefficient $\rho$ between the $S$ and $A$ indexes without considering the flares.
We obtained a correlation coefficient $\rho$ greater than 0.5 for 13 stars in our sample and within this group, and we determined that only seven stars have positive correlation coefficients greater than $\rho > 0.7$ with a confidence level of 95\%. 
These seven stars are shown with red squares in Fig. \ref{s-a_indiv} and they have been identified in the $S-A$ diagram of Fig. \ref{s-a_indivCorr}.
If we consider the flares in our analysis, only PMJ06293-0248 reaches a value of $\rho > 0.7$.

In Table \ref{tab_Halpha_resultados} we show all the results obtained in our study for each M dwarf.
Each star has a series of spectroscopic observations with sampling times extending over 5000 days, with the exception of some stars extracted from the \textit{CONCH-SHELL} catalogue \citep{Gaidos14}.
In our analysis we found that the $S$ and $A$ indexes of stars with intermediate to lower chromospheric emission levels exhibit a higher correlation coefficient ($\rho \geq 0.7$) in our sample.
This result extends what was observed by \cite{GomesdaSilva11}, where they found high correlations only in very active stars.

\begin{figure*}[htb!]
\centering
   \subfigure[\label{rho-logR}]{\includegraphics[width=0.32\textwidth]{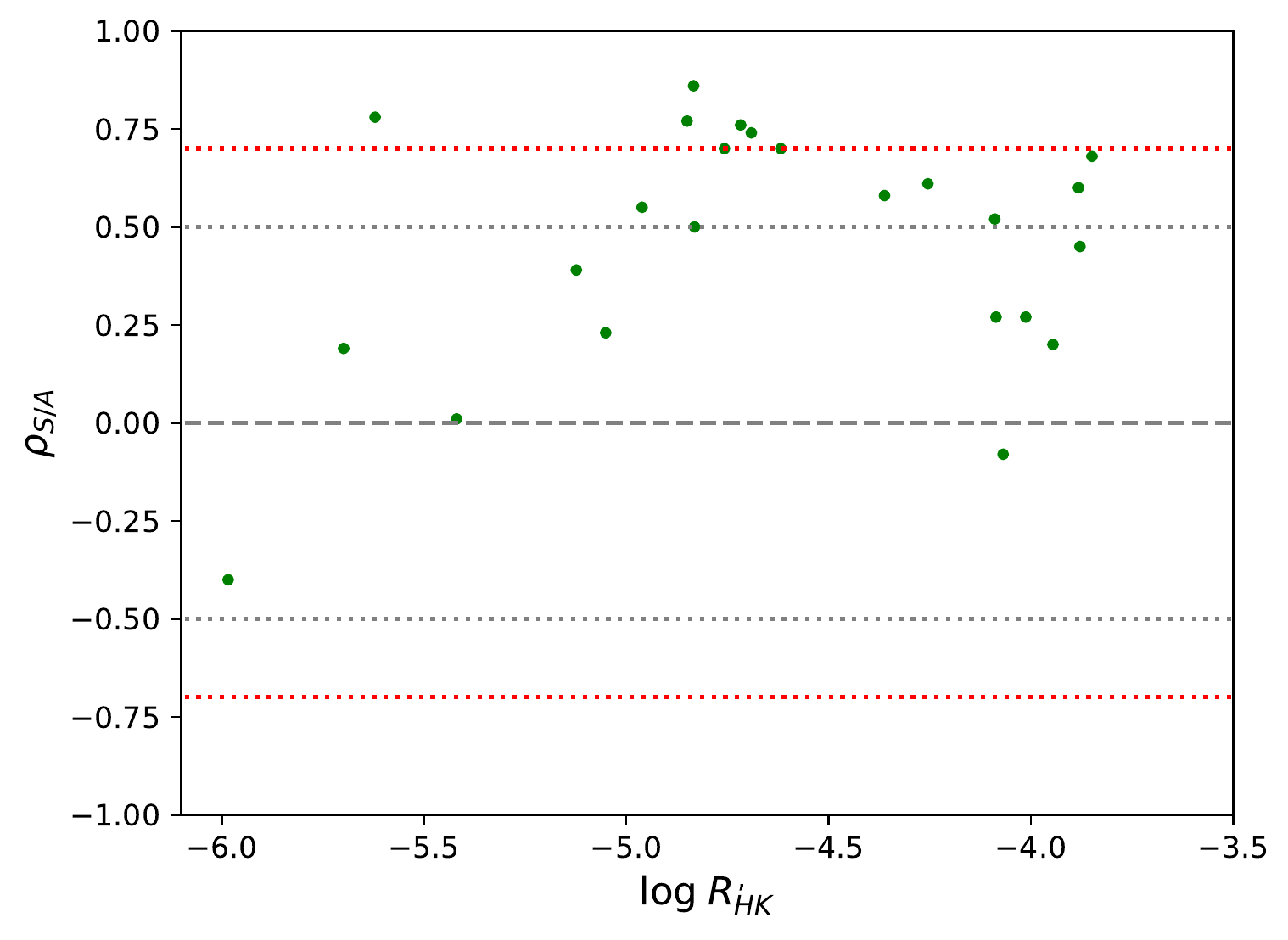}}\hfill
   \subfigure[\label{rho-a}]{\includegraphics[width=0.32\textwidth]{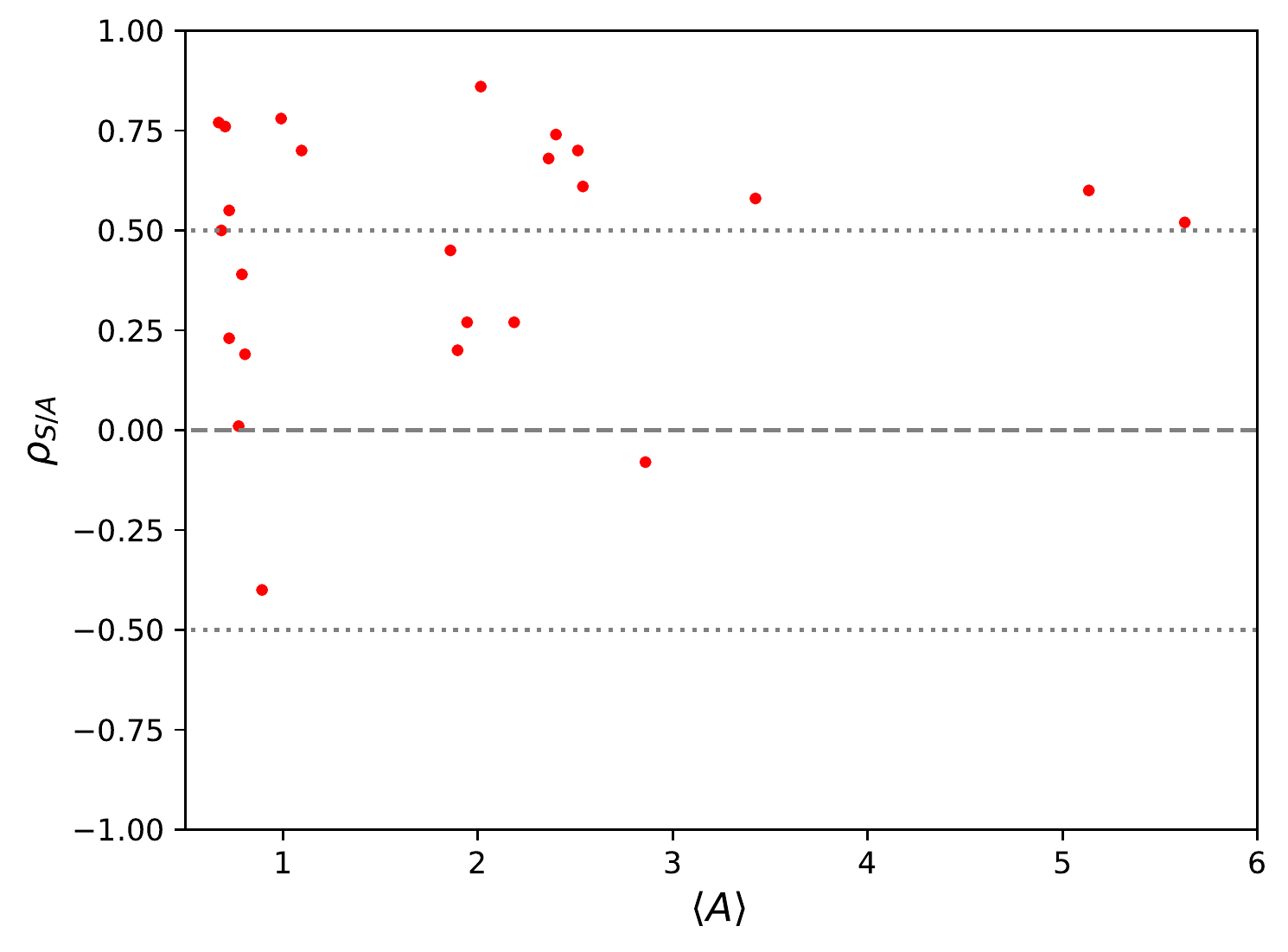}}\hfill
   \subfigure[\label{rho-s}]{\includegraphics[width=0.32\textwidth]{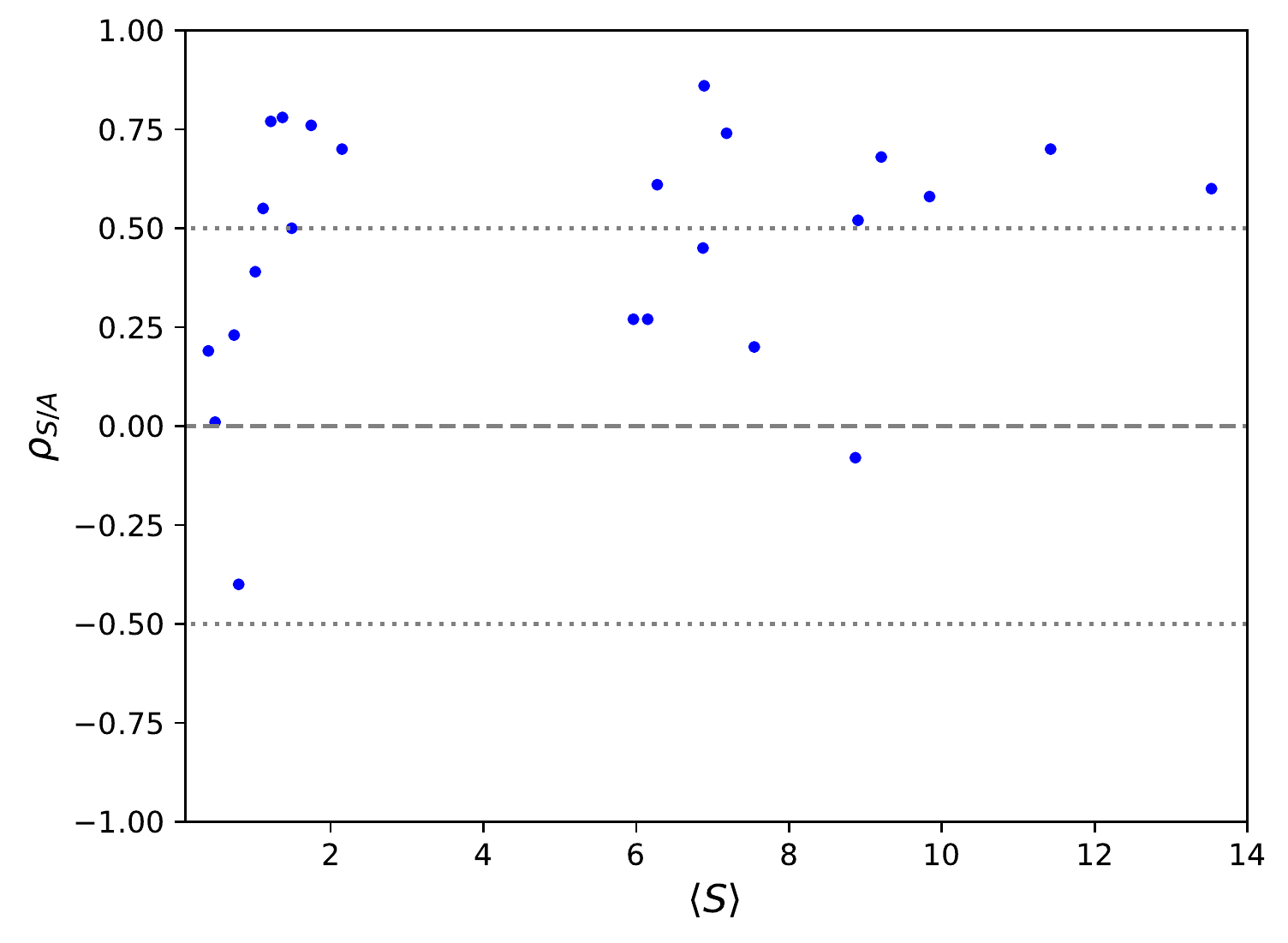}}
   \caption{Correlation coefficient between the $S$ and $A$ indexes ($\rho_{S/A}$) in function of: chromospheric emission level (a), mean value of the $A$ index (b), and  $S$ index for each individual star (c).}
\end{figure*}

However, the vast majority of M dwarfs are positively correlated regardless of activity levels (see Fig. \ref{rho-logR}).
These correlations seem to be independent of the values of the $S$ and $A$ indexes, as shown in Figures \ref{rho-a} and \ref{rho-s}. 
In our sample, we did not find significant ($\rho < -0.5$) anti-correlations between both indexes.
Recently, \cite{Gomesdasilva22} found that an anti-correlation could became a correlation between both indexes when the bandwidth on H$\alpha$ is modified. 
This bandwidth dependence was only studied for solar-type stars (with H$\alpha$ in absorption) employing HARPS spectra. 
In this work, we employed observations from different spectrographs with different resolutions for M stars with both H$\alpha$ in emission and absorption. 
From our study we found that the H$\alpha$ line is very sensitive to stellar rotation and flare activity, so that decreasing the bandwidth of this line to compute the $A$ index for M dwarfs could mean a misinterpretation of the phenomena involved in the high chromosphere where H$\alpha$ is generated.
We finally conclude that the correlation between the indexes for the individual stars is mostly positive throughout the sampling time series.


\section{Summary and conclusion}

In the present work, we have analysed the simultaneous measurements of Ca \scriptsize{II}\normalsize\- and H$\alpha$ indexes obtained for 1796 stellar spectra with broad wavelength coverage. 
We found that indexes associated with flare events do not follow the general trend with a random position in the $S - A$ diagram of Fig. \ref{s-a_crudos}. 
Therefore, flares do not necessarily imply an intensification in both indexes.  
For example, for very active stars such as AD Leo and Proxima Centauri, we can observe broadening in the Balmer lines; this is not observed in the Ca \scriptsize{II}\normalsize\- lines. 
This fact proves that such events  should be monitored in different spectral lines to complement long-term activity analysis based on the Ca \scriptsize{II}\normalsize\- lines.

We found an overall linear trend for the whole sample. 
This linear relation was not observed when we performed the analysis for each individual star.
Seven stars of the sample studied exhibit large correlations with $\rho_{S/A} > 0.7$ regardless of whether the H$\alpha$ line is in emission or absorption, regardless of their chromospheric emission levels and their rotation periods.

On the other hand, we found that there are stars in the sample that deviate from the overall linear trend with large $A$ indexes. 
These stars are characterised by rotation periods shorter than 4 days, large flare rates, and they are located within $3\sigma$ of the saturation regime in the $\log R'_{HK} - P_{rot}$ diagram. 
This implies that fast-rotator flare stars show higher H$\alpha$ emission levels than slower rotators with lower flare activity. 
Low-energy flares are able to produce variation in H$\alpha$ and enhance its emission on active stars \citep{Medina22}. 
This fact could explain the outlier stars coloured in the $S - A$ diagram in Fig. \ref{s-a_estrellas}, where low-energy high-frequency flares could be responsible for the high H$\alpha$ emission. 
Nevertheless, those flares may pass undetected in photometric light curves due to different observational and instrumental reasons (e.g. cadence, instrumental noise, etc.). 

This new study on  the relation between  H$\alpha$ and Ca \scriptsize{II}\normalsize\- activity indexes invite us again to focus on low-mass fast-rotator M stars in stellar activity analysis and, ultimately, in the dynamo theory (\citealt{Newton17, Boudreaux22}). 
In this sense, \citet{Reiners22}  found that Ca \scriptsize{II}\normalsize\- emission saturates at  magnetic fields of around 800G, while H$\alpha$ emission  grows further with stronger fields in fast-rotator M stars.

The latter highlights the importance of having observations with a broad spectral coverage allowing one to analyse  different  indexes of  chromospheric activity simultaneously, not only to discriminate between transient events, but also to analyse the activity at different heights of the stellar atmosphere for a large and diverse set of M dwarfs.

\bibliographystyle{aa}
\bibliography{biblio}

\clearpage
\onecolumn

\begin{landscape}
\tiny
\begin{longtable}{lcccccccccccccc}
\caption{Results obtained throughout the present work. In Col. 7 and 8, we show the mean values of the $S$ and $A$ indexes, respectively, without considering the flare events. In Col. 9 we show the values of the correlation coefficient $\rho$ between indexes with its error in Col. 10. In Col. 11 we show the 95\% confidence levels estimated by the Monte Carlo correlation analysis. In Col. 12 and 13, we show the emission level $\log R'_{HK}$ calculated and the number of observations employed in the present work and, finally, in Col. 14 we indicate the rotation period for each star detected in this work and collected from the literature.}
\label{tab_Halpha_resultados}\\
\hline\hline\noalign{\smallskip}
Star            & SpT   & $B-V$   & T$_{eff}$    & M*            & R*            & $\langle S \rangle$                & $\langle A \rangle$                & $\rho_{S-A}$         & Std. Err.            & 95\% CI& $\log R'_{HK}$ & $N_{obs}$ & $P_{rot}$ & Ref. $P_{rot}$\\
                    &        &         & {[}K{]}      & {[}M*/M$_\odot${]} & {[}R*/R$_\odot${]} &           &            &&&    &    &     & {[}d{]} \\
\hline\noalign{\smallskip}
Gl1                 & dM2    & 1.46  & 3589 $\pm$ 88   & 0.45          & 0.44          & 0.399         & 0.805         & 0.19  & 0.12  & --              & -5.698  $\pm$ 0.239 & 66 & 60.1 & SM15    \\
Gl1049              & dM0e   & 1.443 & --      & --            & --            & 5.963         & 1.944         & 0.27  & 0.20  & --              & -4.086  $\pm$ 0.036 & 18 & 2.634 $\pm$ 0.001 & $\dagger$   \\
Gl1264              & dM2e   & 1.428 & 3228 $\pm$ 92   & 0.61          & 0.57          & 7.545         & 1.896         & 0.20  & 0.17  & --              & -3.946  $\pm$ 0.073 & 28 & 6.562 $\pm$ 0.007 & $\dagger$  \\
Gl182               & dM0e   & 1.373 & 3843 $\pm$ 60   & 0.58          & 0.54          & 6.874         & 1.859         & 0.45  & 0.13  & (0.20 - 0.68)   & -3.879  $\pm$ 0.060 & 39 & 4.414 & K07  \\
Gl388               & dM4e   & 1.3   & 3390 $\pm$ 19   & 0.42          & 0.39          & 8.869         & 2.860         & -0.08 & 0.11  & --              & -4.069  $\pm$ 0.063 & 89 & 2.23  & K12  \\
Gl431               & dM3.5e & 1.547 & 3692 $\pm$ 37   & 0.31          & 0.32          & 8.903         & 5.627         & 0.52  & 0.18  & (0.16 - 0.82)   & -4.089  $\pm$ 0.089 & 14 & 1.0730 $\pm$ 0.0002 & $\dagger$  \\
Gl447               & dM4    & 1.752 & 3192 $\pm$ 60   & 0.168         & 0.1967        & 1.370         & 0.991         & 0.78  & 0.06  & (0.63 - 0.87)   & -5.621  $\pm$ 0.121 & 170 & 121   & B18 \\
Gl479               & dM3e   & 1.535 & 3491 $\pm$ 102   & 0.39          & 0.38          & 2.149         & 1.096         & 0.70  & 0.05  & (0.59 - 0.80)   & -4.757  $\pm$ 0.064 & 90 & 22.5  & SM15    \\
Gl494               & dM0    & 1.483 & 3725 $\pm$ 65    & 0.53          & 0.5           & 6.150         & 2.186         & 0.27  & 0.20  & --              & -4.013  $\pm$ 0.051 & 16 & 2.886 & K12  \\
Gl526               & dM2    & 1.394 & 3521 $\pm$ 56   & 0.47          & 0.47          & 0.737         & 0.724         & 0.23  & 0.13  & --              & -5.051  $\pm$ 0.065 & 54 & 52.3  & SM15  \\
Gl536               & dM1    & 1.47  & 3685 $\pm$ 68   & 0.52          & 0.5           & 1.116         & 0.724         & 0.55  & 0.05  & (0.45 - 0.64)   & -4.961  $\pm$ 0.064 & 188 & 43.9  & SM17  \\
Gl551               & dM5e   & 1.82  & 3050 $\pm$ 100   & 0.1221        & 0.1542        & 11.424        & 2.513         & 0.70  & 0.05  & (0.60 - 0.78)   & -4.618  $\pm$ 0.200 & 111 & 82.9  & K07  \\
Gl699               & dM4    & 1.729 & 3278 $\pm$ 51   & 0.163         & 0.178         & 0.797         & 0.893         & -0.40 & 0.14  & ((-0.67) - (-0.12))  & -5.984  $\pm$ 0.290 & 31 & 145 & TP19    \\
Gl729               & dM4e   & 1.76  & 3213 $\pm$ 60   & 0.14          & 0.19          & 7.183         & 2.401         & 0.74  & 0.05  & (0.65 - 0.84)    & -4.691  $\pm$ 0.118 & 86 & 2.848  & I20 \\
Gl735               & dM3e   & 1.543 & 3431 $\pm$ 86   & 0.34          & 0.34          & 0.85          & 2.539         & 0.61  & 0.09  & (0.42 - 0.77)    & -4.255  $\pm$ 0.084 & 47 & --     & -- \\
Gl803               & dM1e   & 1.423 & 3700 $\pm$ 100   & 0.5           & 0.75          & 9.208         & 2.363         & 0.68  & 0.07  & (0.55 - 0.80)    & -3.849  $\pm$ 0.047 & 64 & 4.85   & I19 \\
Gl825               & dM1    & 1.41  & 3776 $\pm$ 80      & 0.56          & 0.52          & 1.217         & 0.671         & 0.77  & 0.07  & (0.64 - 0.89)    & -4.849  $\pm$ 0.054 & 39 & 40     & B89 \\
Gl908               & dM1    & 1.434 & 3570 $\pm$ 68   & 0.39          & 0.39          & 0.488         & 0.773         & 0.01  & 0.16  & --               & -5.419  $\pm$ 0.112 & 38 & 50.0   & L21 \\
HD180617            & dM3    & 1.515 & 3557 $\pm$ 62   & 0.45          & 0.453         & 1.014         & 0.790         & 0.39  & 0.05  & --               & -5.123  $\pm$ 0.075 & 206 & 46.04  & DA19 \\
HD36395             & dM1.5  & 1.475 & 3612 $\pm$ 57   & 0.6           & 0.58          & 1.745         & 0.704         & 0.76  & 0.04  & (0.67 - 0.84)    & -4.717  $\pm$ 0.091 & 81 & 34     & DA19 \\
HD42581             & dM1    & 1.482 & 3564 $\pm$ --   & 0.58          & 0.54          & 1.491         & 0.684         & 0.50  & 0.08  & (0.34 - 0.65)    & -4.829  $\pm$ 0.069 & 84 & 27.3   & SM16 \\
GJ54.1              & dM5e   & 1.811 & 3062 $\pm$ 62   & 0.14          & 0.19          & 6.889         & 2.015         & 0.86  & 0.02  & (0.82 - 0.89)    & -4.834  $\pm$ 0.220 & 169 & 69.2   & DA19 \\
PMJ09449-1220       & dM6e   & 1.824 & 3025 $\pm$ 62   & 0.14          & 0.19          & --            & --            & --    & --    & --               & -4.284  $\pm$ 0.000 & 3 & 0.44179 $\pm$ 0.00002  & $\dagger$ \\
PMJ02227-6022       & dM5e   & 1.537 & 3380 $\pm$ 72   & 0.29          & 0.3           & --            & --            & --    & --    & --               & -4.279  $\pm$ 0.082 & 5 & 1.1557 $\pm$ 0.0003   & $\dagger$\\
PMJ22449-3315       & dM4e   & 1.516 & 3290 $\pm$ 88   & 0.2           & 0.23          & 13.530        & 5.135         & 0.60  & 0.13  & (0.37 - 0.83)    & -3.883  $\pm$ 0.065 & 28 & 2.356 $\pm$ 0.001   & $\dagger$  \\
PMJ20418-3226       & dM4e   & 1.45  & 3310 $\pm$ 79   & 0.22          & 0.25          &  --            & --            & --    & --    & --              & -3.667  $\pm$ 0.045 & 4 & 1.1943 $\pm$ 0.0005 & $\dagger$  \\
PMJ06293-0248       & dM4.5  & 1.693 & 3168 $\pm$ 40   & 0.14          & 0.19          & 9.840         & 3.425         & 0.58  & 0.15  & (0.28 - 0.84)    & -4.362  $\pm$ 0.148 & 21 & 1.581 & $\dagger$  \\
PMJ06048-3433       & dM6e   & 1.49  & 3079 $\pm$ 62   & 0.14          & 0.19          &  --            & --            & --    & --    & --              & -3.688  $\pm$ 0.000 & 3 & 1.0139 $\pm$ 0.0001 & $\dagger$ \\
PMJ23320-3917       & dM3e   & 1.53  & 3418 $\pm$ 62   & 0.32          & 0.33          &  --            & --            & --    & --    & --              & -4.337  $\pm$ 0.052 & 4 & 3.474 $\pm$ 0.001 & $\dagger$  \\
\hline
\end{longtable}
\tiny
{ $\dagger$ {This work}; (B89) \citet{Byrne89}; (K07) \citet{Kiraga07}; (K12) \citet{Kiraga12}; (G14) \citet{Gaidos14}; (SM15) \citet{SuarezMascareno15}; (SM16) \citet{SuarezMascareno16}; (SM17) \citet{SuarezMascareno17}; (B18) \citet{Bonfils18}; (Gaia18) \citet{Gaia18}; (DA19) \citet{DiezAlonso19}; (I19) \citet{Ibanez18}; (TP) \citet{Toledo19}; (I20) \citet{Ibanez20}; (L21) \citet{Lafarga21}. \par}
\end{landscape}


\begin{appendix}
\section{Flare activity}\label{ap.flare}

To study the magnetic variability of the stars in our study, we employed high-quality photometry obtained by the \textit{Transiting Exoplanet Survey Satellite} (TESS) mission.
The TESS database provides light curves with 20-second and 2-minute cadence for main sequence stars \citep{Ricker14}, which constitute a great basis for  detecting flare-like events (e.g. \citealt{Gunther20,Gilbert22}).  

In order to detect these transient events in the TESS light curve, we analysed the time series with  the FLAre deTection With Ransac Method
(FLATW'RM) algorithm  based on machine-learning techniques \citep{Vida18}\footnote{FLATW'RM is available at \textsf{https://github.com/vidakris/flatwrm/}.}. 
In particular, the FLATW'RM code first determines the stellar rotation period,  and after subtracting the  fitted rotational modulation from the light curve, it detects the flare-like events and reports the  starting and ending time, and the time of maximum flare flux. 

In this section we show the flare activity study for those stars that deviate from the linear trend in Fig. \ref{s-a_estrellas}.
We use blue points to show the light curve of each individual star and red points to show the transient event detected with the FLATW'RM code.
In Table \ref{tab.flares} we summarise our results for each star. 
In Col. 2 we report the rotation periods detected from the GLS periodogram (\citealt{Zechmeister09}) for the non-flaring light curves.
We show in Col. 3 the number of the transient event detected with the FLATW'RM code and, in Col. 4 and 5, we indicate the amplitude (maximum flare flux) and the FWHM of the flare fit.

\begin{table}[h!]
    \centering
    \resizebox{\columnwidth}{!}{%
    \begin{tabular}{lcccc}
    \hline\hline\noalign{\smallskip}
         Star & $P_{rot}$ & \# flares & Amplitude & FWHM  \\
              & [days]    & \# / days &           & [days]\\
    \hline\noalign{\smallskip}
         Gl 431 &  1.0730 $\pm$ 0.0002 & 61/25 & 0.001 - 0.5 & 0.0005 - 1.2\\
         PMJ22449-3315 & 2.356 $\pm$ 0.001 & 15/22 & 0.01 - 0.7 & 0.0009 – 0.01\\
         PMJ02222-6022 & 1.1557 $\pm$ 0.0003 & 40/26 & 0.006 – 0.5 & 0.001 - 0.1\\
         PMJ23320-3917 & 3.474 $\pm$ 0.001 & 29/27 & 0.002 – 0.06 & 0.002 – 0.04\\
         PMJ09449-1220 & 0.44179 $\pm$ 0.00002 & 34/25 & 0.006 – 0.4 & 0.0004 – 0.04\\
         PMJ20418-3226 & 1.1943 $\pm$ 0.0005 & 29/28 & 0.001 – 1.6 & 0.0004 – 0.6\\
         PMJ06048-3433 & 1.0139 $\pm$ 0.0001 & 36/26 & 0.004 – 0.2 & 0.001 – 0.3\\
    \hline
    \end{tabular}
    }
    \caption{Flare activity results obtained in this work for those stars that deviate from the linear trend in Fig. \ref{s-a_estrellas}.}
    \label{tab.flares}
\end{table}

\begin{figure}[htb!]
\centering
   \subfigure[\label{gl431}]{\includegraphics[height=3cm]{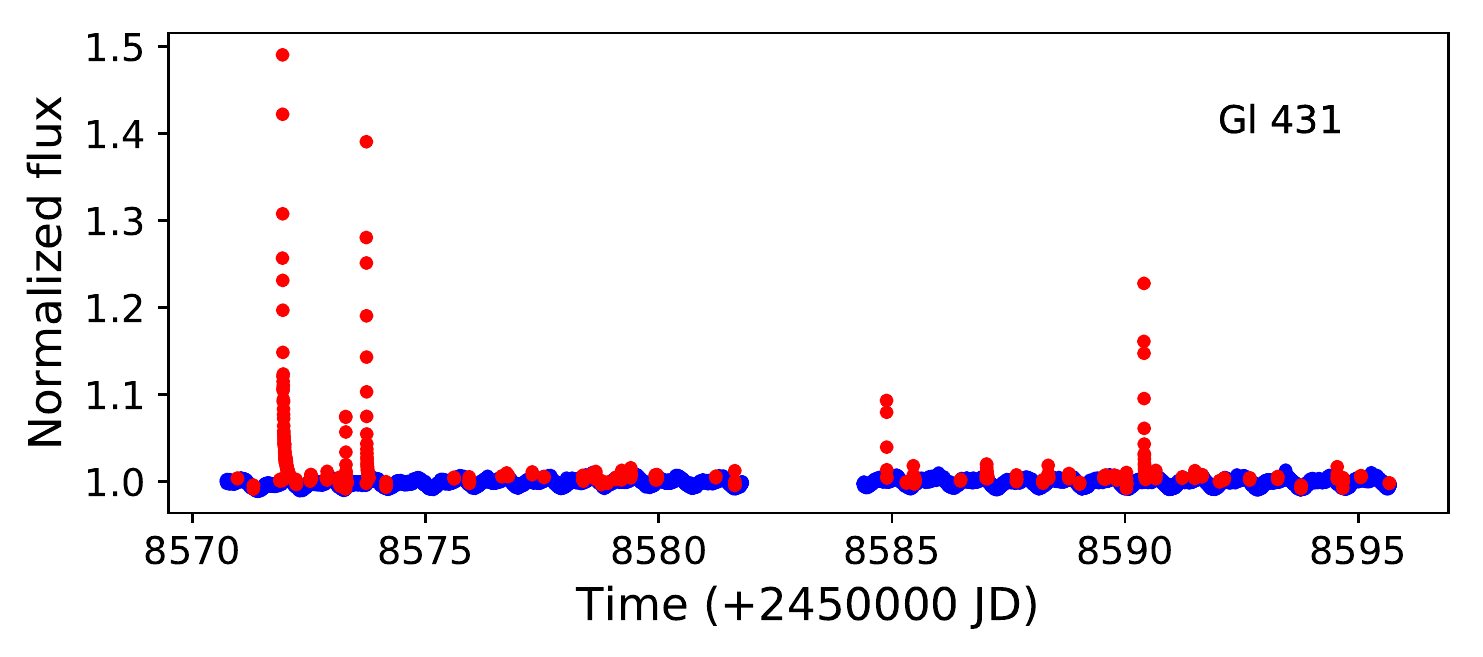}}
   \subfigure[\label{pmj22449}]{\includegraphics[height=3cm]{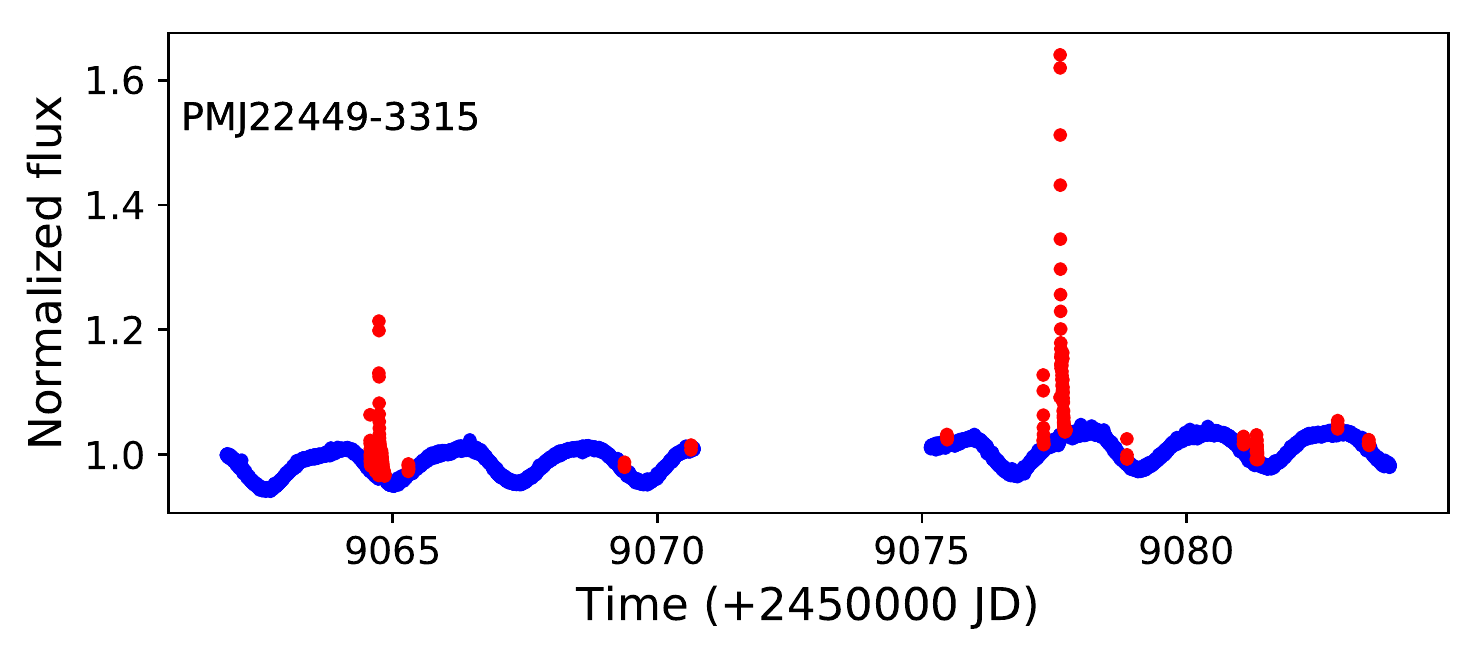}}
   \subfigure[\label{pmj02227}]{\includegraphics[height=3cm]{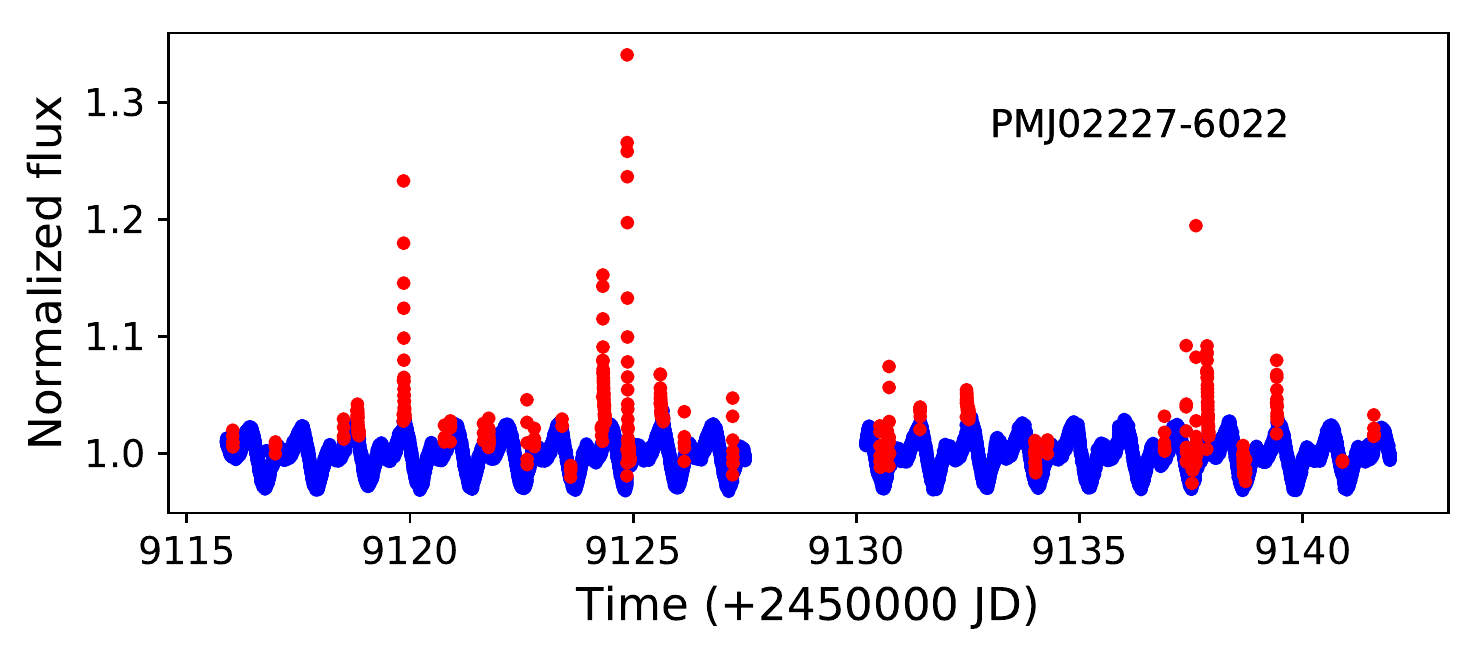}}
   \subfigure[\label{pmj23320}]{\includegraphics[height=3cm]{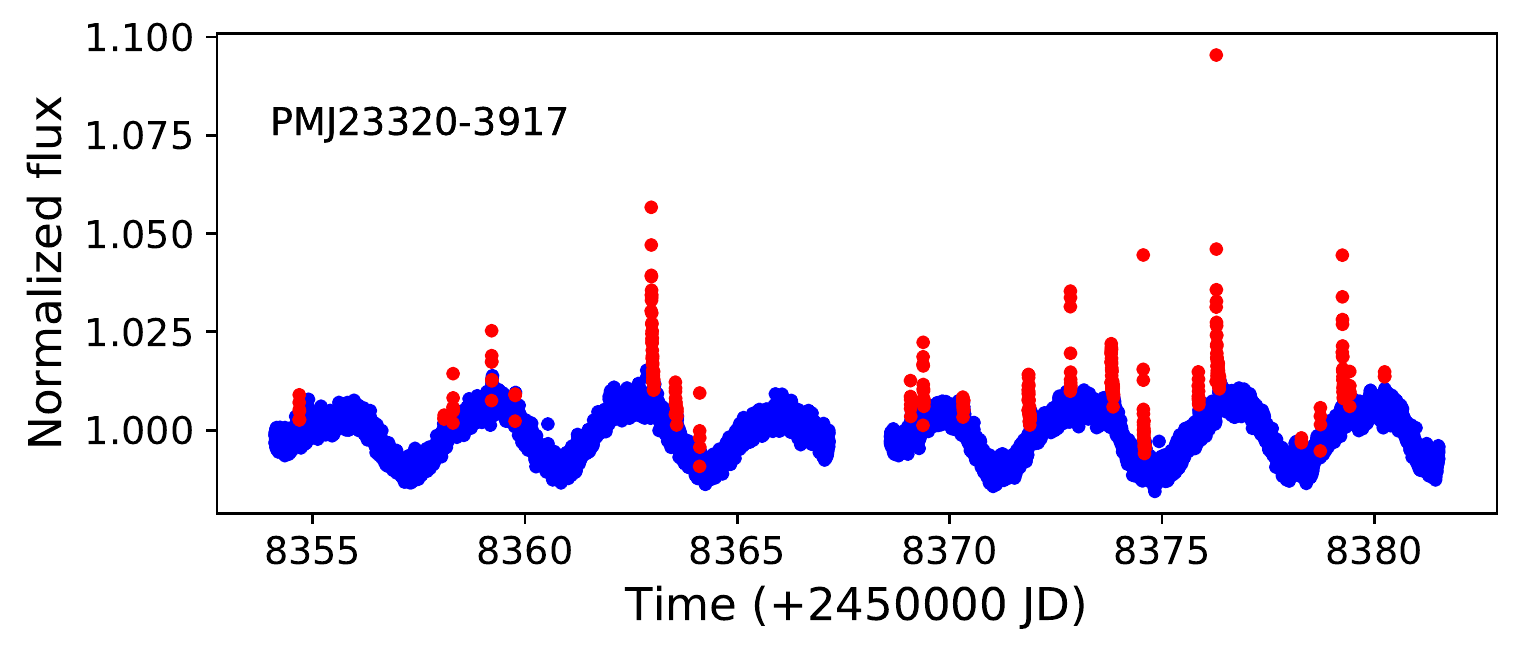}}
   \subfigure[\label{pmj09449}]{\includegraphics[height=3cm]{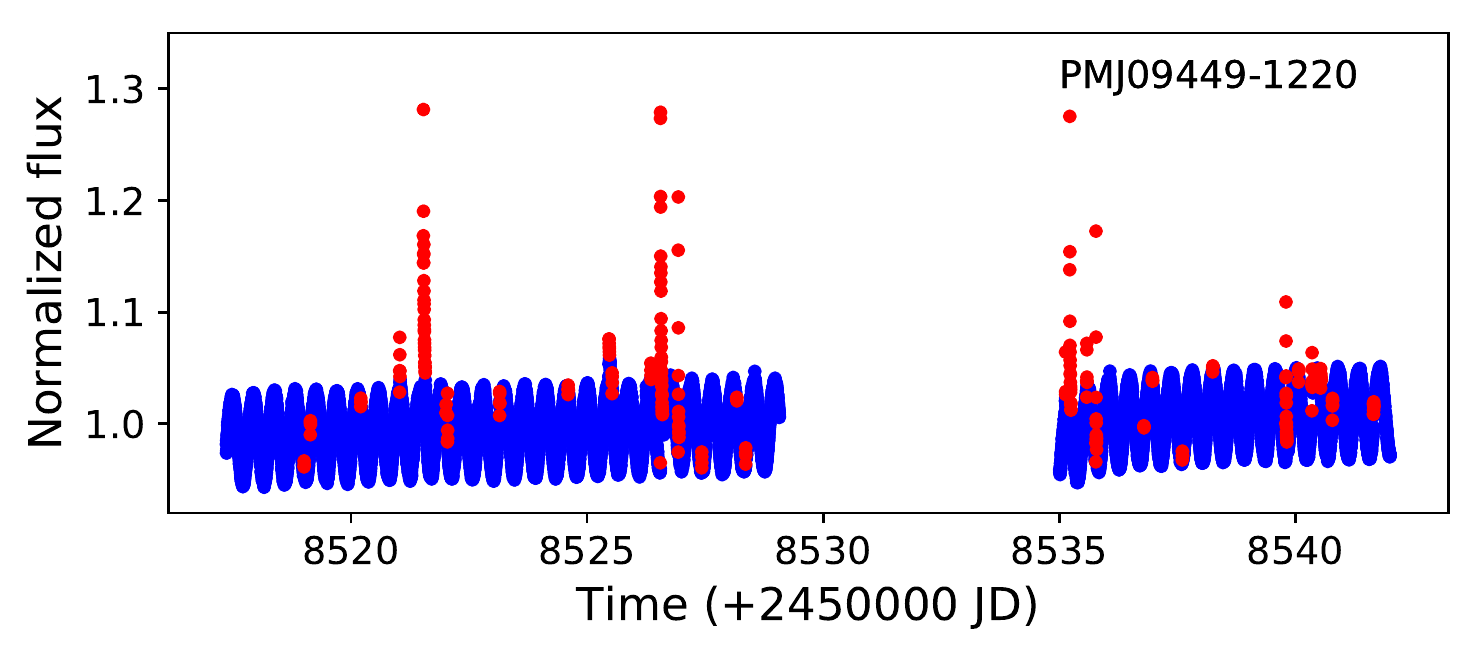}}
   \subfigure[\label{pmj20418}]{\includegraphics[height=3cm]{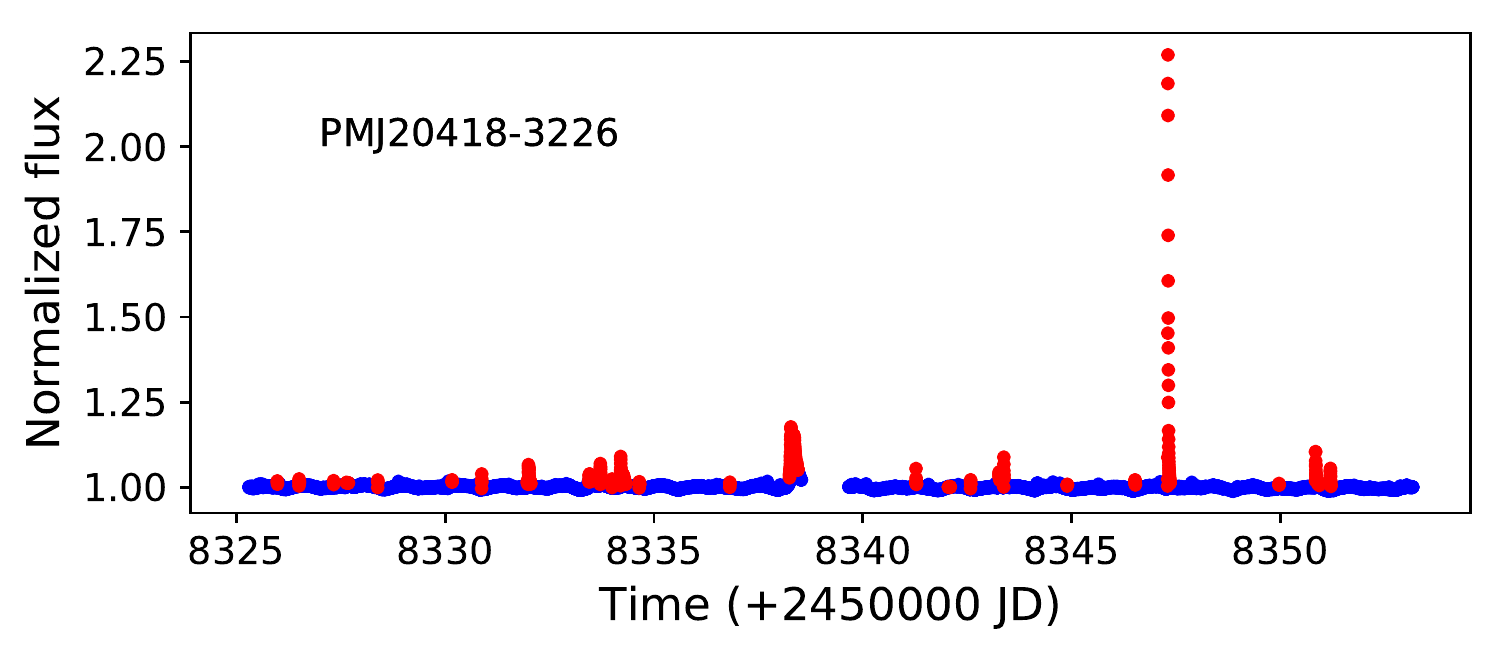}}
   \subfigure[\label{pmj06048}]{\includegraphics[height=3cm]{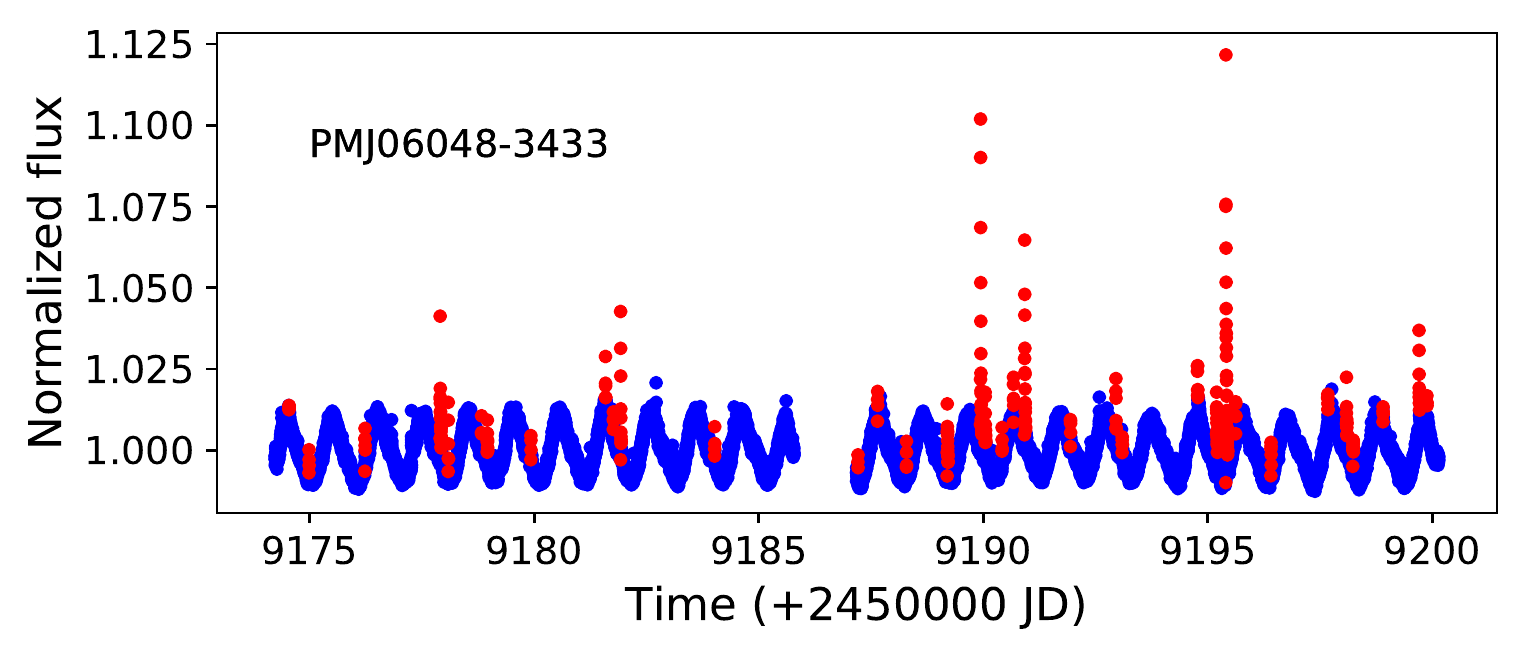}}
   \caption{Flare activity for the seven stars that deviate from the linear trend in Fig. \ref{s-a_estrellas}. We use blue points to show the light curve of each individual star and red points to show the transient event detected with the FLATW'RM code.
   }
\end{figure}

\end{appendix}

\end{document}